\documentclass[superscriptaddress,reprint,amsmath]{revtex4-2}
\usepackage[utf8]{inputenc}
\usepackage[T1]{fontenc}

\usepackage{graphicx} % Required for inserting images

\usepackage[colorlinks=true, allcolors=blue]{hyperref}
\def\equationautorefname#1#2\null{Eq.#1(#2\null)}

\usepackage{xcolor}
\usepackage{xspace}
\usepackage{ulem}

\begin{abstract}
Polaritons are usually described within single-mode cavity QED models. However, nanophotonic environments typically involve several modes that spectrally overlap and interfere, giving rise to sharp dip features such as Fano profiles in the electromagnetic spectral density. Here, we identify these features as interference-induced resonances, effective electromagnetic modes with complex, non-Hermitian couplings to quantum emitters. We show that these modes hybridize with emitters to form polaritons even when the system parameters do not satisfy the single-mode strong-coupling criterion. Moreover, the resulting polaritons differ in their decay rates, a phenomenon we term imaginary Rabi splitting. Extending the analysis to ensembles, we find that coupling to interference-induced resonances produces long-lived polaritons that can outlast excitonic dark states. Numerical simulations of a realistic hybrid metallodielectric platform confirm these predictions and demonstrate their robustness against disorder and loss. Our results reveal a new polaritonic regime beyond the single-mode description, offering new opportunities for controlling light-matter interactions in complex electromagnetic environments.
\end{abstract}

\begin{document}
\author{Anael Ben-Asher}
\email{anaelba@tauex.tau.ac.il}
\affiliation{Departamento de Física Teórica de la Materia Condensada and Condensed Matter Physics Center (IFIMAC), Universidad Autónoma de Madrid, E28049 Madrid, Spain}
\affiliation{Department of Physical Chemistry, School of Chemistry, Tel Aviv University, Tel Aviv 6997801, Israel}

\author{Antonio I. Fern\'andez-Dom\'inguez}
\email{a.fernandez-dominguez@uam.es}
\affiliation{Departamento de Física Teórica de la Materia Condensada and Condensed Matter Physics Center (IFIMAC), Universidad Autónoma de Madrid, E28049 Madrid, Spain}

\author{Johannes Feist}
\email{johannes.feist@uam.es}
\affiliation{Departamento de Física Teórica de la Materia Condensada and Condensed Matter Physics Center (IFIMAC), Universidad Autónoma de Madrid, E28049 Madrid, Spain}

\title{Interference-induced cavity resonances and imaginary Rabi splitting}
\maketitle

\section{Introduction}
Controlling and understanding light-matter interaction is a central topic in science and technology. In recent years, much attention has been devoted to the regime of strong coupling, where the interaction between quantum emitters (QEs) and confined electromagnetic (EM) modes gives rise to hybrid light-matter states, known as polaritons~\cite{basov2020polariton,garcia2021manipulating}. These inherit both photonic and material properties, and often exhibit collective behavior and high delocalization, stemming from the coherent coupling among multiple emitters~\cite{lidzey1999room,cortese2017collective,perez2024collective}. Furthermore, polaritons can feature high robustness to various sources of noise, disorder, and decoherence~\cite{chavez2021disorder,cohn2022vibrational,baghdad2023spectral,wanasinghe2024motional,chng2024mechanism,schwennicke2024extracting}. As such, they offer exciting opportunities for engineering energy flow~\cite{schachenmayer2015cavity,feist2015extraordinary,zhong2017energy,balasubrahmaniyam2023enhanced,sokolovskii2023multi,sandik2024cavity}, modifying chemical reactivity~\cite{feist2018polaritonic,hertzog2019strong,herrera2020molecular,li2022molecular,Ebbesen2023-fd,Bhuyan2023-se}, and realizing quantum technologies~\cite{hennessy2007quantum,sanvitto2016road,toninelli2021single,kavokin2022polariton}.

The description of light-matter interaction typically relies on simplified cavity QED models involving a single EM mode~\cite{jaynes2005comparison,haroche2006exploring}. The polariton formation is then primarily determined by the coupling strength between this mode and the QEs. When this strength exceeds both cavity and emitter dissipation rates, the system enters the aforementioned strong coupling regime, and two polaritons, separated by an energy gap known as the Rabi splitting~\cite{rabi1937space,weisbuch1992observation}, are formed.
However, recent nanophotonic platforms, such as plasmonic nanostructures~\cite{chikkaraddy2016single,zhang2017sub,kongsuwan2018suppressed,baumberg2019extreme}, metamaterials~\cite{shalaev2007optical,torma2014strong,gandman2017two,lindel2025close}, and hybrid metallodielectric cavities~\cite{barth2010nanoassembled,luo2015chip,cui2015hybrid,son2024strong} require descriptions beyond the single-mode picture. They feature complex, highly structured spectral densities, which encode the strength of QE-photon interactions. In particular, when the EM environment is highly non-Lorentzian~\cite{sauvan2013theory,franke2019quantization,pellegrino2020non}, multiple spectrally-overlapping modes with different decay rates may contribute (and interfere) in the light-matter coupling dynamics that take place when a QE is placed in their vicinity. 
In these environments, polariton formation is no longer determined solely by the standard strong coupling criteria, giving rise to regimes where polaritonic properties differ from the single-mode case, and opening new avenues for controlling the hybridization of photonic and material states. 

In this work, we use the recently developed few-mode quantization approach~\cite{medina2021few} to investigate polaritonic phenomena in complex EM environments. A striking feature in their spectral density is the appearance of sharp dips~\cite{sauvan2013theory,franke2019quantization,pellegrino2020non}.
These are often referred to as Fano resonances~\cite{fano1961effects}, electromagnetically induced transparency (EIT)-like resonances~\cite{harris1990nonlinear,peng2014and},  or antiresonances~\cite{taylor1969antiresonance,nitzan1974photon}, depending on the context and the observable under consideration~\cite{limonov2017fano}. 
Previous cavity QED studies have investigated these spectral features  in different contexts: as intrinsic properties of the EM environment uncoupled to emitters~\cite{limonov2017fano,caselli2018generalized,cao2020fano},  as structures observed in optical spectra that include emitter effects~\cite{rice1996cavity,sames2014antiresonance,plankensteiner2017cavity,Cuartero2021distortion}, or as system-specific realizations of coupled photon-QE setups~\cite{gurlek2018manipulation,lu2021plasmonic,lu2022unveiling,lu2024enhanced}. Here, aiming to shed light into the phenomenology previously reported and to provide a general, unifying description of the interaction of QEs with EM environments exhibiting sharp spectral features, we introduce the term interference-induced resonance. This allows us to describe on the same footing resonances and antiresonances, characterized by purely real- and imaginary-valued light-matter couplings, respectively, and producing symmetric Lorentzian-like maxima and minima in the spectral density. Naturally, highly-asymmetric Fano-like profiles (presenting both spectral dips and peaks) occur in EM environments yielding complex-valued coupling strengths. We show that interference-induced resonances, in general, and antiresonances, in particular,  can play a central role in shaping the properties of single emitters and QE ensembles. We reveal a robust and tunable mechanism for antiresonance-QE hybridization, which gives rise to polaritonic states even when the system parameters do not fulfill the single-mode strong-coupling criteria.
Furthermore, we show that these polaritonic states, being degenerate in energy, present significantly different linewidths, or decay rates. We term this phenomenon imaginary Rabi splitting.

We first analytically describe the coupling between a single QE and its EM environment featuring a dip in the spectral density. We identify the parameters of the interference-induced resonance from its spectral features and demonstrate how the non-Hermitian, complex-valued light-matter coupling can give rise to imaginary Rabi splitting, deriving the decay rates and Hopfield coefficients of the resulting polaritons. We then extend our analysis to a QE ensemble interacting with a single antiresonance, where collective coupling leads to the emergence of a polaritonic state with a longer lifetime than the dark, purely excitonic, QE states. Finally, we numerically illustrate our findings in a hybrid platform that includes a microcavity and multiple QE-plasmon pairs. This allows us to demonstrate how antiresonance-mediated polariton phenomena emerge beyond idealized models, even in the presence of significant disorder.
Our results unveil a new mechanism for polariton formation in  complex, highly structured EM environments. They open new directions for engineering decay rates, controlling exciton population dynamics and collectivity, complementing the existing literature on non-Hermitian physics in nanophotonic systems~\cite{chang2014parity,chen2017exceptional,ben2023non,meng2024strong,kkedziora2024non}.

The paper is structured as follows. In \autoref{sec:theory}, we introduce the theoretical framework describing emitter-antiresonance coupling, and analyze both single- and multi-emitter configurations. In \autoref{sec:results}, we present numerical simulations for a hybrid cavity platform hosting multiple emitters, which confirms the predictions of our analytical model. \autoref{sec:conclusions} summarizes the conclusions of our work.% and discusses future directions.

\section{Theory}\label{sec:theory}
\subsection{Few-mode quantization approach}

To rigorously model highly structured, generally non-Lorentzian EM environments, we use a recently developed few-mode quantization scheme~\cite{medina2021few}. In contrast to the intractable Hilbert spaces inherent to macroscopic QED~\cite{huttner1992quantization,scheel2008macroscopic}, this method  provides a compact description for light-matter interactions in arbitrary nanophotonic systems. Characterized by their spectral density, $\text{J}_{\text{EM}}(\omega)$, these can be accurately represented in terms of a small number of effective bosonic modes, with ladder operators $a_n, a_n^\dagger$, each linearly coupled to the QE. The resulting polaritonic quantum dynamics is described by the Lindblad master equation
\begin{eqnarray}\label{lin}
\dot{\rho} = -i\left[\hat{H}_{\text{tot}},\rho\right] + \sum_{n=1}^M \kappa_n L_{a_n}[\rho],
\end{eqnarray}
where
\begin{eqnarray}\label{Heq}
\hat{H}_{\text{tot}} &=& \hat{H}_e + \sum_{n=1}^M \left( \omega_n a_n^\dagger a_n + D_e g_n (a_n^\dagger + a_n) \right)\nonumber\\ &&+ \sum_{m<n} \omega_{nm}(a_n^\dagger a_m + a_m^\dagger a_n).
\end{eqnarray}
Here, $\rho$ is the system density matrix, including the QE and the bosonic EM modes, $\hat{H}_e$ and $D_e$ are the emitter Hamiltonian and dipole operator, respectively, and $L_{a_n}[\rho] = a_n \rho a_n^\dagger - \frac{1}{2} \{ a_n^\dagger a_n, \rho \}$ is a Lindblad super-operator.
The frequencies and losses of the optical modes, given by $\omega_{n} = \omega_{nm}\delta_{nm}$ and $\kappa_n$, respectively, as well as the interaction strengths between them, $\omega_{nm}=\omega_{mn}$, and between each mode and the QE, $g_n$, are obtained by fitting $\text{J}_{\text{EM}}(\omega)$ to the model spectral density
\begin{equation} \label{Jmod}
\text{J}_{\text{mod}}(\omega) = \frac{1}{\pi} \mathrm{Im} \left\{ \vec{g}^T \frac{1}{\mathbf{H}_{\text{mod}} - \omega} \vec{g} \right\},
\end{equation}
where $\vec{g} = { g_1, g_2, \dots, g_M }$ and the entries of the non-Hermitian matrix $\mathbf{H}_{\text{mod}}$ are $\{\mathbf{H}_{\text{mod}}\}_{nm} = \omega_{nm} - \frac{i}{2} \kappa_n \delta_{nm}$.
The accuracy of the few-mode model is determined by the ability of $\text{J}_{\text{mod}}(\omega)$ to reproduce $\text{J}_{\text{EM}}(\omega)$ over the relevant frequency range. In practice, a small number of modes is often sufficient to achieve an accurate fit using \autoref{Jmod}. Once the parameters are extracted, Eqs.~\eqref{lin} and \eqref{Heq} enable an efficient, numerically exact calculation of QE-photon dynamics, accounting for non-Markovian effects and strong coupling phenomena beyond the single-mode picture.

The extension of the few-mode quantization approach to systems comprising multiple emitters was introduced in Ref.~\cite{sanchez2022few}. For a set of $N$ emitters, the EM environment is characterized by a spectral density tensor $\textbf{J}_{\text{EM}}(\omega)$, with element $\alpha\beta$ describing the environment-mediated coupling between QEs $\alpha$ and $\beta$.
The parameters $g_{n,\alpha}$, $\omega_n$, $\omega_{nm}$, and $\kappa_n$ (note that the latter are independent of the QE index) are obtained by fitting the entries of $\textbf{J}_{\text{EM}}(\omega)$ to
\begin{equation}\label{Jmn}
\text{J}_{\alpha\beta}^{\text{mod}}(\omega) = \frac{1}{\pi} \mathrm{Im} \left\{ \vec{g}_\alpha^T \frac{1}{\mathbf{H}_{\text{mod}} - \omega} \vec{g}_\beta \right\},
\end{equation}
where $\vec{g}_\alpha =\{g_{1,\alpha}, g_{2,\alpha}, \dots, g_{M,\alpha}\}$ collects the coupling strengths of emitter $\alpha$ to the optical modes. 
Then, the Hamiltonian describing the unitary dynamics in a master equation of the form of \autoref{lin} for the QE ensemble is
\begin{eqnarray}
\hat{H}_{\text{tot}}^N &=& \hat{H}_e^N + \sum_{n=1}^M \left( \omega_n a_n^\dagger a_n + \sum_{\alpha=1}^{N} D_e^\alpha g_{n,\alpha} (a_n^\dagger + a_n) \right)\nonumber\\ &&+ \sum_{m<n} \omega_{nm}(a_n^\dagger a_m + a_n^\dagger a_m).
\end{eqnarray}
where $\hat{H}_e^N$ is the multi-emitter Hamiltonian and $D_e^\alpha$ is the dipole operator of emitter $\alpha$.

\subsection{Spectral signatures of interference-induced resonances} \label{SecB}

\begin{figure}[tb]
%\begin{center}
   \includegraphics[width=0.9\columnwidth, angle=0,scale=1,
draft=false,clip=true,keepaspectratio=true]{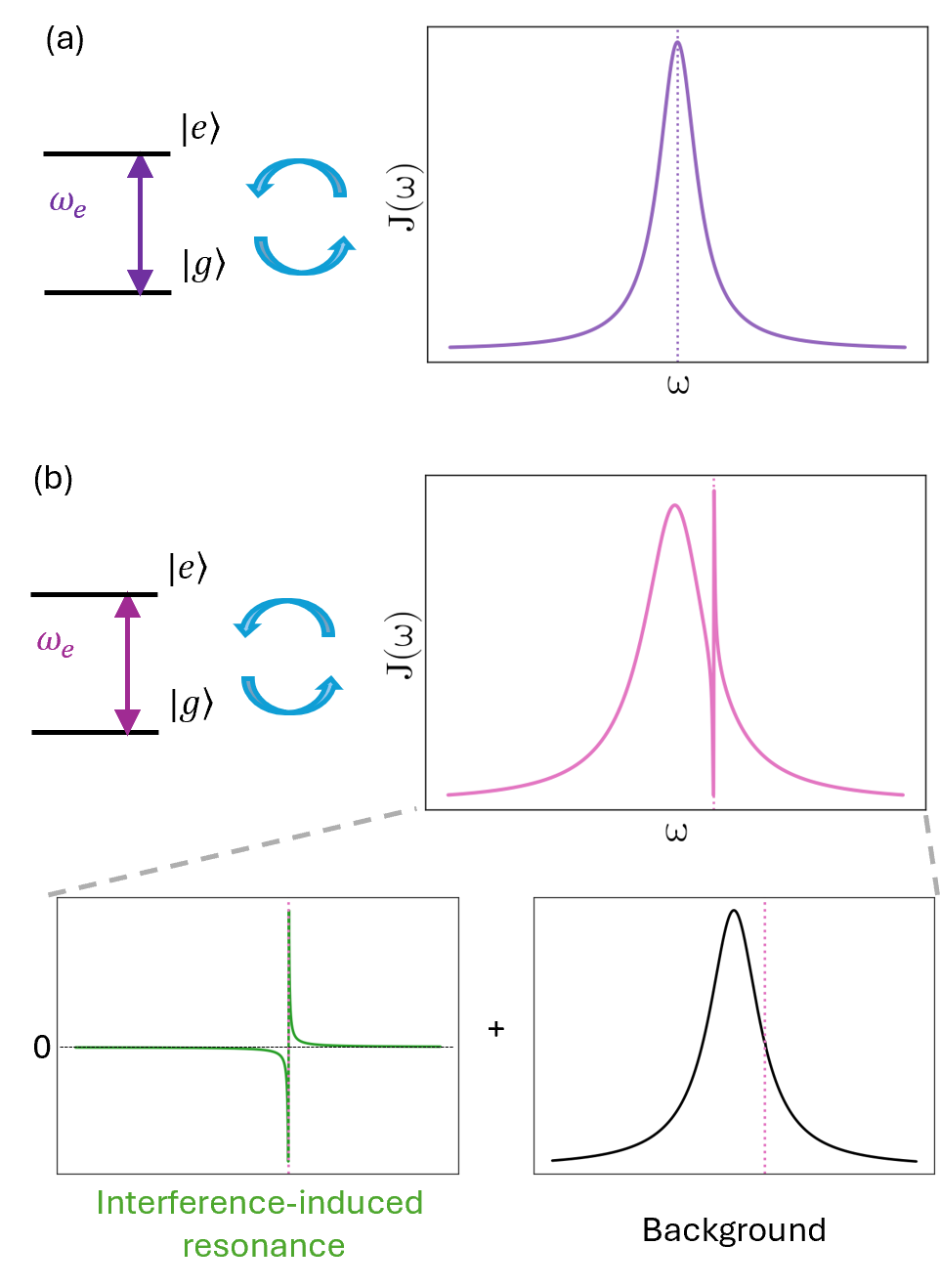}
\caption[]{The coupling of a two-level emitter to (a) a single-mode resonance with a Lorentzian spectral density, and (b) a two-mode cavity with a more complex spectral density. The latter can be described as the sum of a background plus an interference-induced resonance. The vertical dashed lines represent the emitter frequency. In panel (a), the QE frequency coincides with maximum of the Lorentzian-shaped ${\rm J}(\omega)$, whereas in panel (b), ${\rm J}(\omega_e)$ is the minimum of the spectral density.}
\label{fig:1}
%\end{center}
\end{figure}

A key advantage of the few-mode quantization approach is its ability to incorporate modal interactions, $\omega_{nm}$, enabling the description of complex spectral densities. When $\omega_{nm}=0$, $\rm{J}_{\rm mod}(\omega)$ reduces to a sum of independent Lorentzians~\cite{Delga2014quench,Li2016TO}, as that shown in \autoref{fig:1}(a). 
In contrast, when two modes are coherently coupled, ${\omega_{12} \neq 0}$, the real and imaginary parts of $\mathbf{H}_{\text{mod}}$ do not commute, providing the system with a strong non-Hermitian character. This mode-interaction-induced non-Hermiticity gives rise to interference effects between different decay channels in the system. 
%\aba{I'm not sure if the following sentence is not confusing.}\jf{I agree, it's not clear to me what it means.} This becomes evident by writing \autoref{lin} in the basis of the eigenstates of \autoref{Heq}. 
These interferences can lead to sharp dips in  $\rm{J}_{\rm mod}(\omega)$, as illustrated in \autoref{fig:1}(b). Such dips are the fingerprint of the antiresonant character of an EM environment, and play a central role in this work.

To analyze this phenomenon, we consider the minimal case of two interacting modes with $\omega_{12} = d$, whose spectral density reads
\begin{widetext}
\begin{eqnarray}
{\rm J}_{2\text{modes}}(\omega)=\frac{1}{\pi}\frac{\frac{\kappa_2}{2}(g_1d+g_2(\omega_1-\omega))^2+\frac{\kappa_1}{2}(g_2d+g_1(\omega_2-\omega))^2+\frac{\kappa_1\kappa_2^2}{8}g_1^2+\frac{\kappa_2\kappa_1^2}{8}g_2^2}{\bigg((\omega_1-\omega)(\omega_2-\omega)-\frac{\kappa_1\kappa_2}{4}-d^2\bigg)^2+\frac{1}{4}\bigg(\kappa_1(\omega_2-\omega)+\kappa_2(\omega_1-\omega)\bigg)^2}. \label{J2modes}
\end{eqnarray}
\end{widetext}
When one mode is much broader than the other, $\kappa_2 \gg \kappa_1$, and their interaction is much stronger than the geometrical average of their decay rates $ d \gg \sqrt{\kappa_1\kappa_2}$, the first term in the numerator of \autoref{J2modes} dominates. This leads to a Fano-like spectrum of the form 
\begin{equation*}
{\rm J}_{2\text{modes}}(\omega) \approx \frac{2g_2^2}{\pi\kappa_2}\frac{(q+\Omega(\omega))^2}{(1-\Delta(\omega))^2+\Omega(\omega)^2},
\end{equation*}
with 
\begin{align*}
\Omega(\omega) &= \frac{\kappa_2(\omega_1-\omega)}{2d^2},&  q &= -\frac{\kappa_2 g_1}{2 g_2 d}
\end{align*}
and
\begin{equation*}
\Delta(\omega) = \frac{(\omega_1-\omega)(\omega_2-\omega)}{d^2}.
\end{equation*}
When $\Delta(\omega)$ approaches zero, i.e., in the frequency regime where the two modes interfere, the expression reduces to the standard Fano form~\cite{fano1961effects}. A dip appears  for a finite value of $q$ corresponding to the case where the QE coupling of the narrow mode $g_1$ is sufficiently weak compared to the other coupling strengths.

The Fano-like dip in the two-mode spectral density is thus a robust interference effect that arises when three conditions are met: (i) the modes are interacting significantly, (ii) one of them is much broader than the other, and (iii) the broad mode dominates the emitter coupling. Crucially, we show in the following that this dip is not merely a spectral feature, but reflects a mode of the EM environment, an interference-induced resonance, whose complex-valued coupling to QEs effectively cancels light-matter interactions taking place through other EM modes.

\subsection{Interpretation of the Fano lineshape via an effective single-mode model}\label{ar_parameters}

Here we show that the Fano lineshape and the dip in the spectral density discussed above can be interpreted as originating from a single mode of the EM environment, an interference-induced resonance. This is characterized by a complex-valued coupling to the QE, $g_ \mathrm{ir} = g_{R} + i g_{I}$. The spectral density for such a single-mode environment can be constructed using the prescription in \autoref{Jmod}, having
\begin{eqnarray}
\text{J}_ \mathrm{ir}(\omega) = \frac{\frac{\kappa_ \mathrm{ir}}{2}(g_R^2 - g_I^2) - (\omega_ \mathrm{ir} - \omega)g_R g_I}{\pi \left[(\omega_ \mathrm{ir} - \omega)^2 + \frac{\kappa_ \mathrm{ir}^2}{4}\right]},\label{Jii}
\end{eqnarray}
where $\omega_ \mathrm{ir}$ and $\kappa_ \mathrm{ir}$ are the mode frequency and decay rate, respectively. \autoref{Jii} shows clearly the different character of light-matter coupling enabled by the complex character of $g_ \mathrm{ir}$ in interference-induced resonances. For $g_I=0$, the coupling is purely real, and the spectral density acquires a positive, purely Lorentzian shape, the lineshape characteristic of an EM resonance. On the contrary, if $g_R=0$, the coupling is purely imaginary, and ${\rm J}_ \mathrm{ir}(\omega)$ becomes a negative Lorentzian, this is the fingerprint of the antiresonance. In this case, and in any intermediate configuration in which ${\rm J}_ \mathrm{ir}(\omega)$ acquires negative values, the EM environment must involve other modes, since the spectral density of a physical system must remain positive. It can then be expressed as  $\text{J}(\omega) = \text{J}_ \mathrm{ir}(\omega) + \text{J}_{bg}(\omega)$,  where $\text{J}_{bg}(\omega)$ accounts for the background contributions from the other EM modes. This is illustrated in the lower panel of \autoref{fig:1}(b), which shows that the light-matter interaction at the dip of ${\rm J}(\omega)$ results from a near-complete cancellation between the background, with a real-valued coupling, and the interference-induced resonance, with complex coupling $g_ \mathrm{ir}$.

In the following, we present two approaches, one algebraic and one perturbative, to extract the interference-induced resonance parameters (including its complex-valued coupling strength) based on the few-mode quantization scheme. The former involves the direct diagonalization of the Hamiltonian describing the EM environment, which yields both the modal complex frequencies and coupling strengths to a QE. The latter relies on the adiabatic elimination of broad and off-resonant EM modes, resulting in an effective single-mode description of the EM environment.

\subsubsection{Diagonalization of $\mathbf{H}_{\rm{mod}}$}
The interference-induced resonances can be obtained by diagonalizing $\mathbf{H}_{\text{mod}}$ in \autoref{Jmod}, which describes the environment EM modes and their mutual interactions. The eigenvalues correspond to the complex frequencies of the EM eigenmodes, and their QE couplings are given by $\vec{g}^T \vec{\nu}_i$, where $\vec{\nu}_i$ is the $i$-th eigenvector of $\mathbf{H}_{\text{mod}}$ (associated with the $i$-th EM mode). These eigenmodes are closely related to quasinormal modes of the EM environment~\cite{sauvan2013theory, pellegrino2020non}, which can alternatively be obtained by solving Maxwell's equations with outgoing boundary conditions.
In general, all EM eigenmodes are interference-induced resonances, as they exhibit complex-valued light-matter coupling strengths. As demonstrated by \autoref{Jii}, a dip in ${\rm J}_ \mathrm{ir}(\omega)$ arises when $g_I$ dominates over $g_R$ and the mode acquires an antiresonant character. When the remaining modes of the EM environment are weakly coupled to the QE, it is a useful approximation to use a single-mode Hamiltonian for the interaction between this antiresonance-like mode and the emitter, treating the rest as a background, as done, e.g., in Ref.~\cite{sanchez2024mixed}.

We demonstrate here the extraction of the interference-induced parameters by diagonalizing $\mathbf{H}_{\text{mod}}$ for the case of two coupled modes discussed in \autoref{SecB} in the limit of moderate interaction, $\frac{2d}{|\omega_2-\omega_1-i\frac{\kappa_2-\kappa_1}{2}|}\ll1$. This yields the eigenvalues
\begin{equation}
\lambda_{1,2}=\omega_{1,2} - i \frac{\kappa_{1,2}}{2}-\frac{d^2}{\omega_2-\omega_1-i\frac{\kappa_2-\kappa_1}{2}}, \label{lambda}  
\end{equation}
with eigenvectors expressed in the basis of the bare environment modes as 
\begin{align}
\nu_1&=N_1\begin{pmatrix}
1 &\,\, -\frac{d}{\omega_2-\omega_1 - i \frac{\kappa_2-\kappa_1}{2}}
\end{pmatrix}^T, \nonumber \\
\nu_2&=N_2\begin{pmatrix} \frac{d}{\omega_2-\omega_1 - i \frac{\kappa_2-\kappa_1}{2}} &\,\,\,\,\,\, 1 \end{pmatrix}^T. \nonumber
\end{align}
The normalization factors $N_1, N_2$ are approximately valued one. The first eigenmode is mostly dominated by the narrow bare mode ($\kappa_1\ll\kappa_2$), and it is coupled to the QE through 
\begin{equation} \label{gtilde1}
\tilde{g}_1=N_1\left(g_1-\frac{dg_2}{\omega_2-\omega_1-i\frac{\kappa_2-\kappa_1}{2}}\right),
\end{equation}
while the coupling for the second eigenmode, with a large component of the broader bare mode, is 
\begin{equation}
\tilde{g}_2=N_2\left(g_2+\frac{dg_1}{\omega_2-\omega_1-i\frac{\kappa_2-\kappa_1}{2}}\right). \nonumber
\end{equation}
Note that although $g_1$ and $g_2$ are real, the effective couplings above are complex due to interference between the bare modes. In particular, $\tilde{g}_1$ is predominantly imaginary when $g_1$ is small and $|\omega_1-\omega_2|\ll \kappa_2$, so that the imaginary part of the second term is the larger contribution to \autoref{gtilde1}. In this regime, the first eigenmode acts as an antiresonance, such that $\lambda_1=\omega_ \mathrm{ir}-i\frac{\kappa_ \mathrm{ir}}{2}=$ and $g_ \mathrm{ir}=\tilde{g}_1$, producing a dip in the spectral density as in \autoref{Jii}. The second eigenmode, can be treated as a background contribution when its (predominantly real) QE-coupling is within the Markovian regime, $g_2\ll\kappa_2$. 

\subsubsection{Adiabatic elimination of broad and detuned modes}
Another approach for extracting the interference-induced parameters within the few-mode description is through adiabatic elimination. The starting point is the non-Hermitian Hamiltonian obtained by absorbing the anticommutator in the Lindblad super-operators in \autoref{lin} into \autoref{Heq},
\begin{eqnarray}
\hat{H}_{\text{tot}}^{\text{NH}} &=& \hat{H}_e + \sum_{n=1}^M \left( (\omega_n-i\frac{\kappa_n}{2}) a_n^\dagger a_n + D_e g_n (a_n^\dagger + a_n) \right) \nonumber \\ 
&&+ \sum_{n,m<n} \omega_{nm}(a_n^\dagger a_m + a_m^\dagger a_n). 
\end{eqnarray}
We focus on the regime where a Fano dip arises in the spectral density due to destructive interference between a narrow mode and a broad one. In accordance with our previous notation, we denote the single-photon state in the narrow EM mode, whose frequency lies near that of the QE, $\omega_e$, as $a_1^\dagger|0\rangle$, where $|0\rangle$ is the vacuum. Moreover, we assume that all the remaining modes in the EM environment are either broad or far-detuned from the QE and the narrow EM mode, thus satisfying either ${\kappa_n \gg g_n, \omega_{1n}, \kappa_1, |\omega_1 - \omega_e|}$ or ${|\omega_n - \omega_e| \gg g_n, \omega_{1n}, \kappa_1, |\omega_1 - \omega_e|}$ ($n>1$). Transforming $\hat{H}_{\text{tot}}^{\text{NH}}$ into the rotating frame of the narrow mode through the unitary operator $\hat{U} = e^{-i \omega_1 (\sum_{n=1}^M a_n^\dagger a_n) t}$, we can adiabatically eliminate all the modes with $n>1$, assuming that they remain close to their instantaneous steady state. This procedure yields a single EM mode with a shifted complex frequency and a complex QE coupling, corresponding to the interference-induced resonance of the EM environment. Due to the adiabatically-eliminated modes, which act as the EM background described above, the QE frequency is also Lamb-shifted and acquires a dissipative character.

For the case of a two-level emitter with $\hat{H}_e=\omega_e\sigma^+\sigma^-$ and dipole moment operator $D_e=\mu_e(\sigma^+ + \sigma^-)$, and for an EM background  composed of non-interacting modes (i.e., $\omega_{nm}=0$ for $n\neq1$ and $m\neq1$), the adiabatic non-Hermitian Hamiltonian obtained under the rotating-wave approximation reads~\cite{lu2022single,lu2022unveiling}
\begin{multline}\label{Had}
\hat{H}_{ad} = \bigg(\omega_e-\sum_{n=2}^M\frac{\mu_e^2 g_n^2}{\Delta_n}\bigg)\sigma^ + \sigma^- + \bigg(\omega_ \mathrm{ir}-i\frac{\kappa_ \mathrm{ir}}{2}\bigg) a_1^\dagger a_1 \\
+ \mu_e g_ \mathrm{ir} (a_1^\dagger \sigma^- + \sigma^+ a_1),
\end{multline}
where ${\Delta_n=\omega_n-\omega_1-i\frac{\kappa_n-\kappa_1}{2}}$, ${g_ \mathrm{ir}=g_1-\sum_{n=2}^M\frac{\omega_{1n}g_n}{\Delta_n}}$ and $\omega_ \mathrm{ir}-i\frac{\kappa_ \mathrm{ir}}{2}=\omega_1-i\frac{\kappa_1}{2}-\sum_{n=2}^M\frac{\omega_{1n}^2}{\Delta_n}$. 
As anticipated, the QE frequency is shifted and becomes complex (dissipative) due to the EM background. Notably, this shift is identical to the one obtained by treating each adiabatically-eliminated mode as a Markovian bath with Lorentzian-shaped spectral density, and assuming $\omega_1 \approx \omega_e$. \autoref{Had} shows that the parameters of the narrow EM mode, its complex frequency and QE coupling are also renormalized to describe the interference-induced resonance, responsible for the Fano dip in the spectral density (rather than the $n=1$ mode). In the two-mode case ($\omega_{12}=d$), the adiabatic elimination reproduces the same complex frequency as the diagonalization of $\mathbf{H}_{\rm{mod}}$, expanded to leading order in $d$ (\autoref{lambda}), and yields the complex-valued QE coupling given by \autoref{gtilde1} evaluated at $N_1=1$, i.e., in the adiabatic regime where 
$a_1\approx a_ \mathrm{ir}$. This agreement confirms that, within its validity range, the adiabatic elimination approach accurately captures both the effect of the EM background and the emergence of the interference-induced resonance.

\subsection{Imaginary Rabi splitting}
The single-mode description of the interference-induced resonance not only provides a physical interpretation for the Fano lineshape in the spectral density of the EM environment of a QE, but also offers practical tools for analyzing light-matter coupling in the system. In particular, it reveals a regime of an imaginary Rabi splitting, arising from the complex-valued character of the QE coupling and enabling the design of EM decay rates. 
Moreover, this framework reveals that polaritons, hybrid light-matter states, can form even without direct coupling to the narrow mode that dominates the interference-induced resonance, i.e., $g_1 = 0$. We demonstrate these phenomena using the Hamiltonian in \autoref{Had}, considering the case where $M = 2$ and $g_1 = 0$, $\kappa_1\to0$ and $|\omega_e-\omega_1|\to0$.
In this case, $\hat{H}_{ad}$ has two eigenstates in its first excitation manifold 
\begin{eqnarray}\label{pn}
|+\rangle=\frac{1}{\sqrt{d^2+\mu_e^2g_2^2}}(d\sigma^+ -\mu_e g_2 a_1^\dagger)|0\rangle, 
\end{eqnarray}
and
\begin{eqnarray}\label{pb}
   |-\rangle=\frac{1}{\sqrt{d^2+\mu_e^2g_2^2}}(\mu_e g_2\sigma^+ + d a_1^\dagger)|0\rangle.
\end{eqnarray}
Their associated eigenvalues are
\begin{eqnarray}
\epsilon_+=\frac{\omega_e(1-\eta)+(\omega_1-i\frac{\kappa_1}{2})(1+\eta)}{2},\label{en}
\end{eqnarray}
and
\begin{eqnarray}
\epsilon_-=\frac{\omega_e(1+\eta)+(\omega_1-i\frac{\kappa_1}{2})(1-\eta)}{2}-\frac{\mu_e^2g_2^2+d^2}{\Delta_2},\label{eb}
\end{eqnarray}
where $\eta=\frac{\mu_e^2g_2^2-d^2}{\mu_e^2g_2^2+d^2}$ is a dimensionless parameter measuring the relative strength of the interaction of EM mode 2 with the QE and with mode 1. 

As seen from Eqs.~\eqref{pn}-\eqref{pb}, for $\eta = \pm1$, the system eigenstates reduce to the bare emitter and EM modes. In contrast, when $\eta = 0$ ($\mu_e g_2$ and $d$ are comparable), the two eigenstates become maximally hybridized, forming polaritons with QE and EM components. The parameter $\eta$ thus quantifies the degree of hybridization between the QE and the interference-induced resonance. Notably, the polaritonic states emerge even though $g_1 = 0$, meaning there is no direct coupling between the QE and the EM mode governing the interference-induced resonance. Instead, the hybridization arises indirectly through the modal interaction, $d$, to the eliminated mode.
This light-matter coupling mechanism gives rise to a complex splitting between polariton eigenvalues, in contrast to the typically real-valued Rabi splitting in conventional polaritons.
The origin of this complex splitting lies in the EM background, given by mode 2. The eigenstate $|+\rangle$ is decoupled from it, while $|-\rangle$ inherits a complex frequency shift from it. As seen from Eqs.~\eqref{en}-\eqref{eb}, in the limit $\eta \to 0$, or when $\kappa_1 \to 0$ and $|\omega_e - \omega_1| \to 0$, this shift is given by $\frac{\mu_e^2 g_2^2 + d^2}{\omega_2 - \omega_1 - i\kappa_2/2}$. When $|\omega_2 - \omega_1| > \kappa_2$, the real part of this splitting dominates, yielding a real Rabi splitting, as observed in Refs.~\cite{verre2019transition,maciel2023probing} despite the weak direct light-matter coupling. However, when $\omega_2 = \omega_1$, the splitting becomes purely imaginary, and the interference-induced resonance can be considered as an antiresonance.
In this configuration, polariton states present a purely imaginary Rabi splitting due to indirect, interference-mediated hybridization, a non-Hermitian analog of conventional strong coupling. This phenomenon offers a mechanism to modulate polaritonic decay rates in dissipative EM environments, and can also give rise to single-photon emission, as demonstrated in Refs.~\cite{lu2022unveiling,ben2023non}.

\subsection{Coupling multiple emitters to an interference-induced resonance}
As discussed above, interference-induced resonances arise from the destructive overlap between EM modes and hence, unlike conventional cavity modes, they do not necessarily have a localized spatial character. As a result, the collective coupling of a QE ensemble to an interference-induced resonance is not inherently defined, and depends critically on how each emitter interacts with its own, interfering EM environment. In the following, we identify two distinct mechanisms by which such collective coupling can emerge. 

The first mechanism arises from the identical coupling of all the QEs to a shared background mode. This, in turn, interacts with a narrower mode, giving rise to an interference-induced resonance that can, depending on the system parameters, manifest as an antiresonance with purely-imaginary light-matter coupling. This occurs, for example, in hybrid metallodielectric cavities, where a lossy plasmonic resonance, interacting equally with each QE in an ensemble, is coupled to a low-loss dielectric mode. This indistinguishability in the QE-plasmon interaction leads to an adiabatic Hamiltonian that can be expressed solely in terms of a collective bright state of the ensemble~\cite{tavis1969approximate,fink2009dressed,del2015quantum}. This has the form of \autoref{Had} for $M=2$ and with single-emitter operators replaced by bright-state ones.
The coupling of the interference-induced resonance to this collective bright state $g_ \mathrm{ir}$ then scales with the square root of the number of emitters $\sqrt{N}$ due to the collective QE-plasmon coupling (mirroring the scaling of conventional collective, light-matter coupling~\cite{tavis1969approximate,fink2009dressed}). The resulting polaritonic energies follow Eqs.~\eqref{en}-\eqref{eb}, with the substitution $g_2 \to \sqrt{N}g$ where $g$ is the individual QE-plasmon coupling. Apart from the two polaritonic states, the system supports $N-1$ dark, exciton states with energy $\omega_e$, which are decoupled from the EM environment. Notably, in the limit $\kappa_1 \to 0$ and $|\omega_e - \omega_1| \to 0$, the narrow polariton given by \autoref{pn} becomes degenerate with the dark states, and becomes sensitive to perturbations and disorder~\cite{del2015quantum}. 

The second mechanism takes place when each QE is locally coupled to a different background mode, and all of them collectively interact with a shared, narrow EM mode to generate an interference-induced resonance (or an antiresonance).
As sketched in \autoref{fig:2}(a), this can occur in a low-loss Fabry-Perot cavity embedding an array of plasmonic nanoparticles, each coupled to a different QE.
Since the EM environment interacts differently with each emitter, the effective Hamiltonian obtained from the adiabatic elimination of the background modes explicitly involves all of them, rather than only the bright state of the ensemble.
As a result, the dark states in this setup are not fully decoupled from the EM environment, but acquire complex shifts due to the renormalization of the emitter frequencies, $\omega_e - \frac{\mu_e^2 g^2}{\Delta}$, originating from their individual (non-collective) interactions with the EM background. Here, $g$ is the local QE-plasmon coupling and $\Delta$ is the detuning between the complex plasmonic and photonic frequencies. 
Yet, the interference-induced resonance is only coupled to the collective bright state of the ensemble, with the coupling scaling as $\sqrt{N}$. The scaling here arises from the collective interference of multiple background modes (rather than collective light-matter interaction). Therefore, this scaling is also included in the complex frequency of the interference-induced resonance.
The polaritonic eigenvalues again follow Eqs.~\eqref{en}-\eqref{eb}, with the substitution $d \to \sqrt{N} d$ (and $g_2 \to g$) where $d$ is the photon-plasmon coupling strength.
Counterintuitively, the narrow polariton can then become longer-lived than the excitonic dark states, highlighting the nontrivial nature of the collective interference that gives rise to the interference-induced resonance.

In \autoref{sec:results}, we numerically study the second setup, which is particularly relevant for physical implementations based on QE-nanoparticle coupling, where only a few emitters can be placed near each metallic nanoparticle to interact with its plasmonic mode.

\section{Numerical results}\label{sec:results}
\begin{figure}[tb]
%\begin{center}
   \includegraphics[width=0.75\columnwidth, angle=0,scale=1,draft=false,clip=true,keepaspectratio=true]{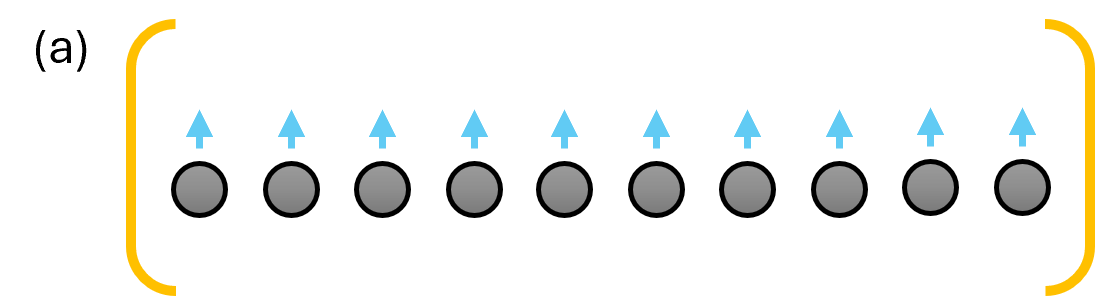}

\vspace{0.4cm}

   \includegraphics[width=0.95\columnwidth, angle=0,scale=1,
draft=false,clip=true,keepaspectratio=true]{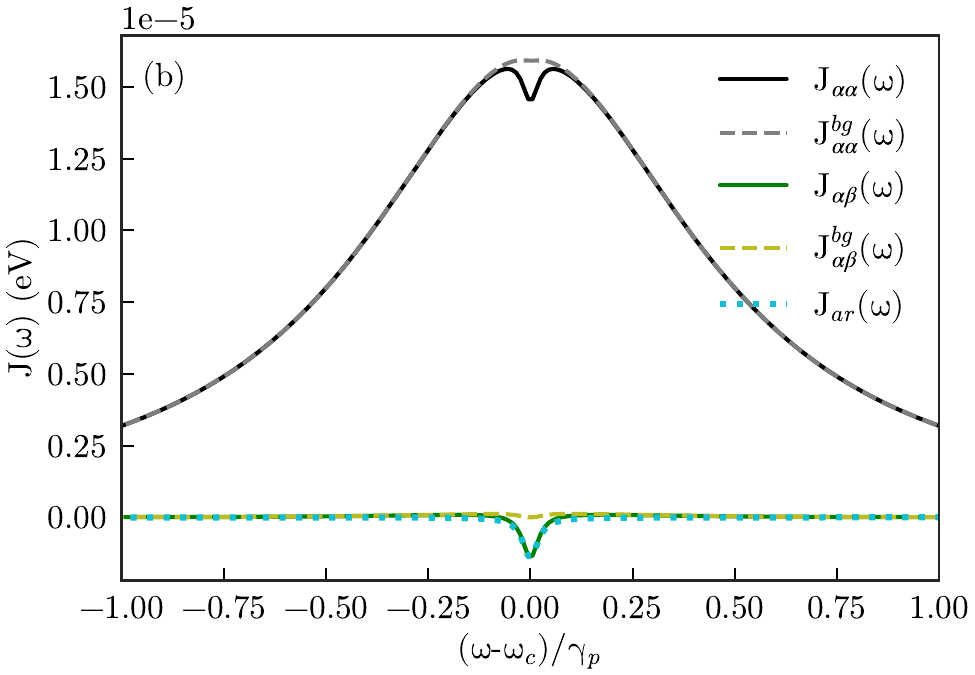}
\caption[]{(a) Scheme of a hybrid cavity consisting of a Fabry-Perot cavity and multiple metal nanoparticles, each interacting with a different QE. (b) The diagonal entries of the spectral density, J$_{\alpha\alpha}(\omega)$, and the cross ones, J$_{\alpha\beta}(\omega)$, for the QE ensemble in panel (a). ${\rm J}_ \mathrm{ar}(\omega)={\rm J}_ \mathrm{ir}(\omega)$ is the spectral density of the antiresonance formed in this setup, and J$_{\alpha\alpha}^{bg}(\omega)$ and J$_{\alpha\beta}^{bg}(\omega)$ are the spectral densities obtained when subtracting J$_ \mathrm{ar}(\omega)$ from J$_{\alpha\alpha}(\omega)$  and J$_{\alpha\beta}(\omega)$, respectively. }
\label{fig:2}
%\end{center}
\end{figure}

To numerically illustrate the imaginary Rabi splitting arising from the coupling of an emitter to an antiresonance, beyond the analytical models presented in \autoref{sec:theory}, we consider an array of $N$ identical QE–nanoparticle pairs placed inside a photonic cavity, as shown in \autoref{fig:2}(a). In this configuration, each nanoparticle supports a plasmonic mode that interacts exclusively with its adjacent QE through a coupling strength $g$, while all plasmonic modes collectively couple to the cavity mode with strength $d$. The direct coupling between the QEs and the cavity is assumed to be negligible, owing to the large mode volume of the photonic mode.

\subsection{Spectral density analysis}
The diagonal and off-diagonal elements of the spectral density tensor, J$_{\alpha\alpha}(\omega)$  and J$_{\alpha\beta}(\omega)$, calculated from \autoref{Jmn} for the system shown in \autoref{fig:2}(a), are plotted in \autoref{fig:2}(b) in solid black and green lines, respectively. These correspond to $N = 10$ QEs, with $g = \frac{\kappa_p}{20}$ and $d = \frac{\kappa_p}{10\sqrt{N}}$, with plasmon loss $\kappa_p = 0.1$ eV and cavity linewidth $\kappa_c = \frac{\kappa_p}{200}$. Both plasmonic and photonic modes are at resonance, $\omega_p = \omega_c$. 
We can observe that $\text{J}_{\alpha\alpha}(\omega)$, which describes the spectral density experienced by each individual emitter, features a broad peak together with the symmetric dip characteristic of the antiresonance. In contrast, as the EM environment renders the QEs distinguishable, the cross-correlated spectral density, $\text{J}_{\alpha\beta}(\omega)$, which captures the EM-mediated interaction between emitters, exhibits only the dip lineshape. We identify the interference-induced contribution, which in this case corresponds to an antiresonance, as $\text{J}_ \mathrm{ar}(\omega)=\text{J}_ \mathrm{ir}(\omega)$ as defined in \autoref{Jii}. This is plotted as a dotted light blue line in \autoref{fig:2}(b), using parameters derived from the adiabatic elimination of the plasmonic modes, which closely match those obtained by diagonalizing $\mathbf{H}_{\text{mod}}$, see~\autoref{ar_parameters}.
The diagonal contribution of the EM background, $\text{J}^{bg}_{\alpha\alpha}(\omega)$, yields the broad Lorentzian profile in dashed gray line, while its off-diagonal counterpart, $\text{J}^{bg}_{\alpha\beta}(\omega)$, in dashed light green line, vanishes. 
These differences in the spectral density components reflect the physical picture discussed above: the EM background couples to each QE independently, and thus affects the exciton dark states. On the contrary, the antiresonance governs the collective dynamics in the system, which is encoded in the polariton states.

\subsection{Decay rates and imaginary Rabi splitting}
\begin{figure}[t]
%\begin{center}
   \includegraphics[width=0.95\columnwidth, angle=0,scale=1,
draft=false,clip=true,keepaspectratio=true]{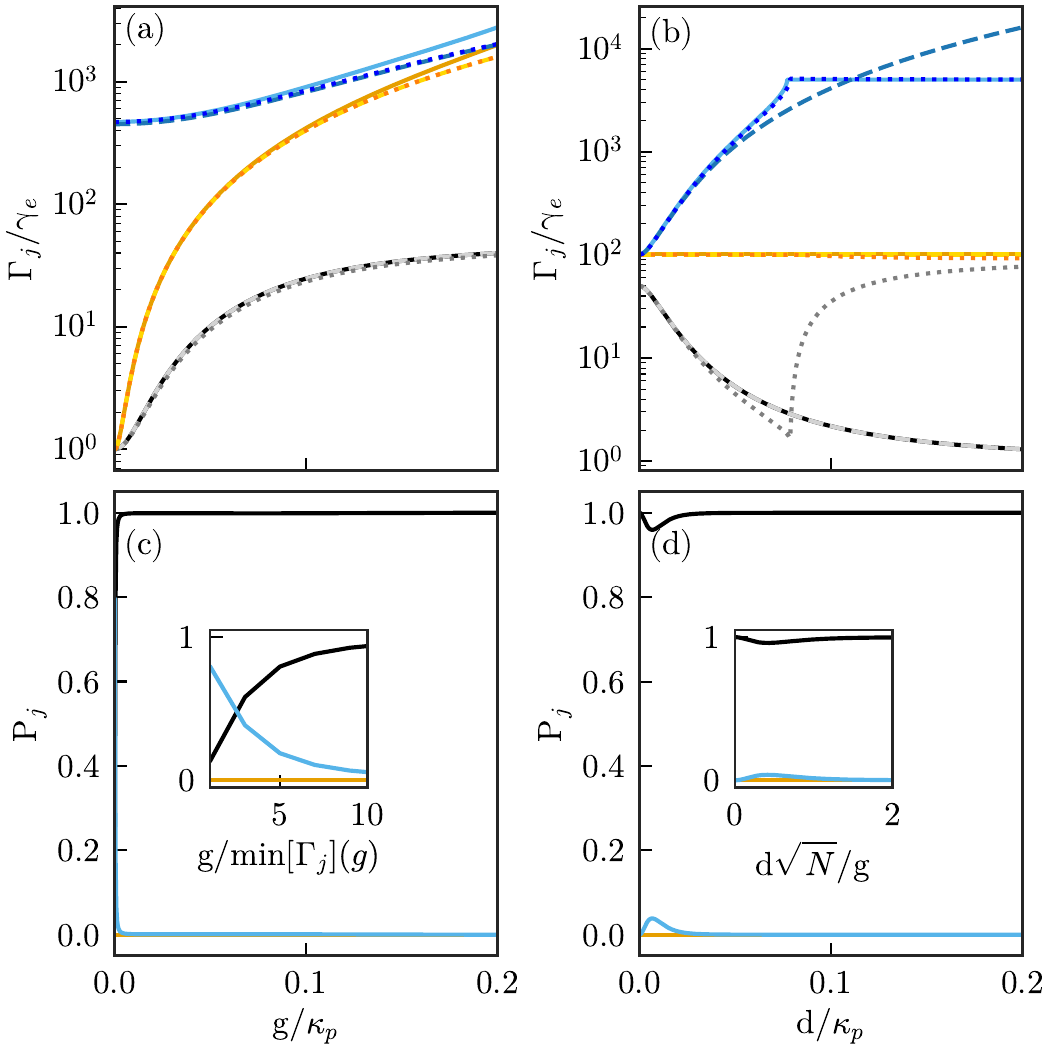}
\caption[]{Decay rates of the eigenstates resulting from the QE-antiresonance coupling in the setup shown in \autoref{fig:2}(a) as a function of: (a)  $g$ when  $d\sqrt{N}=\kappa_p/10$, and (b) $d$ when $g=\kappa_p/20$. 
Blue and black/gray lines correspond to polariton states, while the orange line renders the dark exciton states that do not interact with the antiresonance. Solid lines were obtained from full calculations, while dashed (adiabatic elimination of the plasmonic modes) and dotted (coupling to the antiresonance with the EM background treated in a Markovian manner) lines plot approximate calculations. 
(c) Steady state populations for the system in (a) under coherent driving of the cavity mode, normalized by the total population at the first excited manifold. (d) Same as (c) but for the system in (b). The insets in (c-d) demonstrate the conditions for almost-complete population of the narrow eigenstate.
}
\label{fig:3}
%\end{center}
\end{figure}

In Figs.~\ref{fig:3}(a)-(b), we analyze the decay rates, $\Gamma_j$, of the eigenstates of the system in \autoref{fig:2}(a) as a function of: (a) the QE-plasmon coupling, $g$ ($d \sqrt{N}= \kappa_p / 10$), and (b) the plasmon-cavity coupling, $d$ ($g = \kappa_p / 20$), where the QE frequencies $\omega_e$ are at resonance with the antiresonant dip frequency given by $\omega_ \mathrm{ar}=\omega_c$. These correspond to the imaginary parts of the complex eigenvalues, $\epsilon_j - i\Gamma_j/2$. Note that in the weak coupling regime ($g<\kappa_p/4$ and $d\sqrt{N}<\kappa_p/4$), $\epsilon_j=\omega_e$ for all the eigenstates. The solid lines are obtained by numerically diagonalizing the non-Hermitian Hamiltonian
\begin{multline}
\hat{H}_{\text{tot}}^{N\text{(NH)}} = \tilde{\omega}_c a_c^\dagger a_c + \sum_{\alpha=1}^N \bigg[\tilde{\omega}_p a_{p,\alpha}^{\dagger} a_{p,\alpha} + \tilde{\omega}_e \sigma^+_{\alpha} \sigma^-_{\alpha}\\ + g\left(a_{p,\alpha}^{\dagger}\sigma^-_{\alpha}+\sigma^+_{\alpha}a_{p,\alpha}\right)
+d\left(a_{p,\alpha}^{\dagger} a_c + a_c^\dagger a_{p,\alpha}\right)\bigg],
\end{multline}
where $\tilde{\omega}_c=\omega_c-i\frac{\kappa_c}{2}$, $\tilde{\omega}_p=\omega_p-i\frac{\kappa_p}{2}$ and $\tilde{\omega}_e=\omega_e-i\frac{\gamma_e}{2}$ are the complex frequencies of the cavity, plasmons and emitters, respectively. We set the QE spontaneous decay rate to $\gamma_e=\frac{\kappa_p}{10000}$, while the other parameters are the same as in \autoref{fig:2}. 
Although $\hat{H}_{\text{tot}}^{N\text{(NH)}}$ has $2N+1$ eigenstates in the single-excitation manifold, we plot in Figs.~\ref{fig:3}(a)-(b) only the $N+1$ states with the lowest $\Gamma_j$. These include the two polaritonic states, a broad one (blue) and a narrow one (black), which are hybrids of QE and Fabry-Perot cavity excitations. The $N$ dark states (orange) are composed almost entirely of QE excitations, with only a small plasmonic contribution. The remaining states that have higher decay rates (not shown) are primarily of photonic or plasmonic character.

To establish the link between the system eigenvalues and the antiresonance, we also include in Figs.~\ref{fig:3}(a)-(b) approximate results obtained by considering the coupling of the QEs to a single EM mode, the antiresonance itself, rather than to the full set of photonic and plasmonic modes (see \autoref{ar_parameters}).
The dotted lines correspond to eigenvalues computed from this reduced model, whose parameters are extracted by diagonalizing $\mathbf{H}_{\text{mod}}$, and the remaining EM background is treated perturbatively within the Markovian approximation~\cite{sanchez2024mixed}.
On the other hand, the dashed lines are obtained from an effective Hamiltonian derived by adiabatically eliminating the plasmonic modes. 
Both approximate approaches reproduce the full results within a broad range of $g$ and $d$, enabling interpretation of all their key features via the analytical results discussed in \autoref{sec:theory}.
In particular, Figs.~\ref{fig:3}(a)-(b) demonstrate the imaginary Rabi splitting, i.e., the splitting in decay rates between the two polaritons, arising from the coupling to the antiresonance. 
In addition, the results also reveal the effect of the EM background on the dark states (and their decoupling from the antiresonance). Their linewidth increases with plasmon-QE coupling, $g$, indicating a stronger interaction with the background, but is insensitive to the plasmon-cavity coupling, $d$, which affects only the QE-antiresonance coupling. Note that when $d \gg g$ ($\eta = -1$), the narrow state, given by \autoref{pn} replacing $\sigma^+$ by the collective bright state and $a_1^\dagger\to a_c^\dagger$, $g_2 \to g$, $d \to \sqrt{N} d$, is essentially the bright state of the QE ensemble. On the contrary, for $g \gg d$ ($\eta = 1$), it becomes the bare cavity mode. This explains why the decay rate of the narrow state increases with $g$ but decreases with $d$, as the Fabry-Perot cavity loss is larger than the QE loss ($\kappa_c > \gamma_e$). 
Importantly, the regime $d\gg g$, which is more easily achieved physically owing to the typically stronger plasmon-cavity coupling, naturally leads to a long-lived W-like collective state~\cite{dur2000three}, relevant in quantum information protocols~\cite{yeo2006teleportation}.

Figs.~\ref{fig:3}(a)-(b) also show the breakdown of the two approximate methods above. We can observe that both deviate from the exact results for the decay rate of the dark states and broad polariton when $g$ approaches the regime of strong coupling, see \autoref{fig:3}(a). This is a direct consequence of the failure of the Markovian approximation and the adiabatic elimination for the EM background. The narrow state, which becomes the bare cavity mode in this regime, remains unaffected by this breakdown.
However, when $d$ reaches strong coupling, making the broad state a hybrid plasmonic-photonic mode (degenerate with another hybrid mode of the same kind, not shown), each approximation captures the decay rate of a different state, see \autoref{fig:3}(b).
The adiabatic approximation fails to reproduce $\Gamma_j$ for the broad state, as it assumes different timescales for the Fabry-Perot cavity and plasmon dynamics.
Still, it accurately captures the decay rate of the narrow state, which is mainly the bare bright state. This suggests that at high plasmon-cavity coupling, the full EM environment can be effectively described as a Markovian bath.
In contrast, the antiresonance description obtained by diagonalizing $\mathbf{H}_{\text{mod}}$ successfully captures the decay rate of the broad state, which involves only EM excitations, but fails for the narrow state that also includes QE excitations.
This failure demonstrates that the assumption that a single EM mode dominates the light-matter interaction (used in this reduced model) breaks down in the regime of strong plasmon-cavity hybridization.

\subsection{Steady-state population}
Figs.~\ref{fig:3}(c)-(d) show the steady-state population, $P_j$, of the eigenstates studied in Figs.~\ref{fig:3}(a)-(b) under weak coherent driving of the Fabry-Perot cavity by an external laser. We use the same line colors and parameters as in those panels. The populations are normalized by the total steady-state population within the single-excitation manifold.
In the weak coupling regime ($g<\kappa_p/4$ and $d\sqrt{N}<\kappa_p/4$), all the states are on-resonance with the laser, and their population is proportional to their overlap with the cavity mode (which is the driven one), divided by their decay rate. Accordingly, Figs.~\ref{fig:3}(c)-(d) show that the narrow polariton, which has the smallest $\Gamma_j$, dominates the steady-state population. In particular, it becomes the most populated eigenstate when its cavity component, weighted by $g$, becomes larger than its decay rate, given by $\text{min}[\Gamma_j]$ as a function of $g$, as shown in the inset of \autoref{fig:3}(c). 
The inset of \autoref{fig:3}(d) shows a slight deviation from complete population of the narrow polariton when $d\sqrt{N} < g$, originating from the less pronounced imaginary Rabi splitting in this regime, see \autoref{fig:3}(b). The predominance of the narrow eigenstate in the steady-state population highlights a key advantage of the imaginary Rabi splitting: by engineering decay rates, it enables quantum state preparation under continuous driving. This feature is particularly relevant in the collective regime considered here, where a hybrid polaritonic state, despite its bright character, can be engineered to outlive the dark states and thereby dominate the steady-state dynamics.

\subsection{Effect of detuning and robustness to disorder}
\begin{figure}[tb]
%\begin{center}
   \includegraphics[width=0.95\columnwidth, angle=0,scale=1,
draft=false,clip=true,keepaspectratio=true]{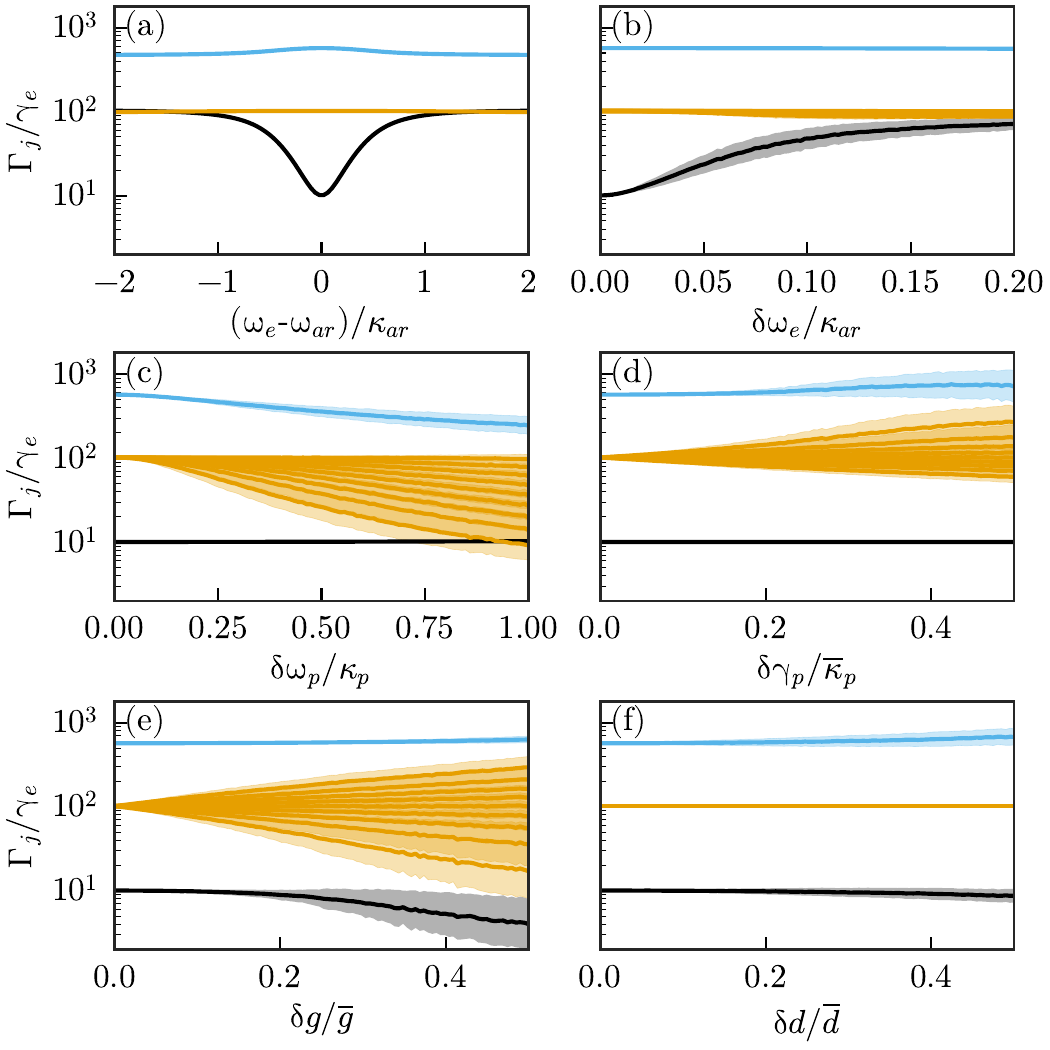}
\caption[]{Decay rate of the eigenstates in the system in \autoref{fig:2}(a) as a function of: (a) the detuning between the QE frequencies and the antiresonance (no disorder); (b) disorder in the QE frequencies; (c) disorder in the plasmon frequencies; (d) disorder in the plasmon linewidths; (e) disorder in the plasmon-emitter couplings; and (f) disorder in the plasmon-cavity couplings. The color code is the same as in \autoref{fig:3}.
In (b)-(f), $\delta X$ denotes the standard deviation of a Gaussian distribution with mean $\overline{X}$, and the shaded regions span one standard deviation above and below the geometric mean of the decay rates (solid lines).
}
\label{fig:4}
%\end{center}
\end{figure}

To further highlight the role of the antiresonance, in ~\autoref{fig:4}(a), we study the effect of detuning the QEs across the frequency range spanned by the antiresonance. Specifically, we present the decay rates of the broad (blue), dark (orange) and narrow (black) states as a function of $(\omega_e-\omega_ \mathrm{ar})/\kappa_ \mathrm{ar}$, where $\omega_ \mathrm{ar}$ (that obeys $\omega_ \mathrm{ar}=\omega_c=\omega_p$) and $\kappa_ \mathrm{ar}=\kappa_1+\frac{4Nd^2}{\kappa_p-\kappa_c}$ are the energy and decay rate of the antiresonance, respectively. We use $d=\frac{\kappa_p}{10\sqrt{N}}$ and $g=\frac{\kappa_p}{20}$, for which the approximate approaches discussed above yield the same values of the antiresonance parameters. All remaining parameters are identical to those in Figs.~\ref{fig:2}~and~\ref{fig:3}. We can observe that the imaginary Rabi splitting emerges only when the QE frequency is within the linewidth of the antiresonance, $|\omega_e-\omega_ \mathrm{ar}|<\kappa_ \mathrm{ar}$. This confirms that the phenomenon is confined to the spectral region where the antiresonance mediates light-matter interaction in the system.

To study the robustness of the imaginary Rabi splitting and the different eigenstates in \autoref{fig:3}, we examine in Figs.~\ref{fig:4}(b)-(f) the modification of their decay rates under various types of static disorder. 
Specifically, we introduce disorder in (b) the emitter frequencies, (c) the plasmon frequencies, (d) the plasmon linewidths, (e) the plasmon-QE coupling strengths, and (f) the plasmon-cavity coupling strengths. In each case, the disordered parameter, $X$, is sampled from a Gaussian distribution with mean value $\overline{X}$ equal to its value in \autoref{fig:4}(a) for $\omega_e=\omega_ \mathrm{ar}$, and standard deviation $\delta X$. The shaded regions span one standard deviation above and below the geometric mean of $\Gamma_j$, computed over $1000$ disorder realizations.
The narrow (black) and broad (blue) polaritonic states generally exhibit considerable robustness to disorder across a wide range of parameters, while the dark states (orange) are consistently more sensitive to it. 
Among the different disorder types, the disorder in the QE frequencies in \autoref{fig:4}(b) has the most significant effect. When the standard deviation $\delta\omega_e$ exceeds roughly  $\frac{\kappa_ \mathrm{ar}}{20}$, the collective interaction of the QE ensemble with the antiresonance breaks down, preventing the formation of a long-lived and robust polaritonic state in the system.
In contrast, the narrow eigenstate exhibits remarkable tolerance to disorder in the plasmonic frequency, \autoref{fig:4}(c), and linewidth, \autoref{fig:4}(d). Its decay rate remains unchanged even when the standard deviation reaches a significant fraction of the mean value of the parameter. This is a direct consequence of the full decoupling of this state from the plasmons, in contrast to the dark states and the broad polariton. The effect of larger disorder in QE-plasmon and cavity-plasmon couplings in Figs.~\ref{fig:4}(e) and \ref{fig:4}(f) on $\Gamma_j$ for the narrow eigenstate, however, is moderate, as these parameters control its photonic and QE components. These results further emphasize the feasibility of our setup, which offers an experimental platform for the realization of imaginary Rabi splitting.

%It's interesting to see that the narrow polariton is almost not affected by the plasmon parameters, because it is completely decoupled from it (while the broader one is affected by the plasmon). However, the parameters g,d,we determine the coefficients of the emitters/cavity in the narrow polariton and therefore affect its width.  

\section{Conclusions}\label{sec:conclusions}
We have introduced a general framework for the description of light-matter interaction in structured electromagnetic environments featuring narrow dips in their spectral density. Using a few-mode quantization approach, we have identified these dips as originating from interference-induced resonances, effective electromagnetic modes with complex-valued coupling strengths to the emitters. This interpretation goes beyond previous descriptions of spectral dips as purely optical features, providing a mode-based picture that directly links them to coherent and dissipative light-matter coupling. 
Furthermore, we have demonstrated how these modes can lead to the formation of hybrid light-matter states, i.e., polaritons, even when the system is far from the conventional strong-coupling regime. 

The analytical treatment of a single emitter coupled to an antiresonance, an interference-induced resonance yielding purely imaginary light-matter coupling, revealed that the non-Hermitian nature of the interaction gives rise to an imaginary Rabi splitting, in which the polaritons are separated in their decay rates rather than in their energies. Depending on system parameters, the same mechanism can also give rise to conventional real Rabi splitting, as observed experimentally in Refs.~\cite{verre2019transition,maciel2023probing}. We extended this analysis to collective configurations where emitter ensembles are coupled either to a common, extended mode or to spatially separated, local modes that interfere through a single photonic cavity. In the latter configuration, the emitters are rendered distinguishable by their environment. Nonetheless, we find that a bright collective state can still couple to the antiresonance and form a long-lived polariton state, while the excitonic, dark states become short-lived. This opens the door to robust collective state engineering in the presence of loss and disorder. Numerical simulations in a realistic hybrid metallodielectric platform confirmed these predictions, and proved the validity of our findings beyond ideal, simplified models.

Our results introduce a new mechanism for polariton engineering  in complex electromagnetic environments and highlight interference-induced resonances and antiresonances as resources to control their decay rates and population dynamics. These insights pave the way for a broad range of polaritonic applications and open new directions for utilizing non-Hermitian spectral features in nanophotonics to enhance coherence and enable robust quantum state preparation.

\section*{Acknowledgments}
We acknowledge financial support by the Spanish Ministerio de Ciencia y Universidades—Agencia Estatal de Investigación through grants PID2021-125894NB-I00, PID2021-126964OB-I00, PID2024-161142NB-I00, PID2024-156077OB-I00, EUR2023-143478, and CEX2023-001316-M through the María de Maeztu program for Units of Excellence in R\&D, and from the European Union’s Horizon Europe research and innovation programme under Grant Agreement No. 101070700 (MIRAQLS).
In addition, this project received funding from the European Union's Horizon 2020 research and innovation programme under the Marie Skłodowska-Curie Grant Agreement No 101034324.

\bibliography{sample}

\end{document}